\documentclass[12pt]{article}

\usepackage{graphicx}

\newcommand{\beao}{\begin{eqnarray*}}
\newcommand{\eeao}{\end{eqnarray*}}
\newcommand{\bea}{\begin{eqnarray}}
\newcommand{\eea}{\end{eqnarray}}
\newcommand{\be}{\begin{equation}}

\newcommand{\ee}{\end{equation}}

\newcommand{\hk}{\hspace{0.1cm}}

\newcommand{\rk}{\right)}
\newcommand{\lk}{\left(}

\newcommand{\sli}{\sum\limits}

\renewcommand{\vec}[1]{\mbox{\boldmath$#1$\unboldmath}}

\newcommand{\vB}{{\vec{B}}}
\newcommand{\vA}{{\vec{A}}}

\begin{document}

\begin{center}
{\Large\bf \boldmath Coulomb
gauge Yang--Mills theory \\ in the Hamiltonian approach\footnote{Invited talk given by H. Reinhardt at the international conference
on ``Selected Problems in Theoretical Physics, Dubna 23-27 June
2008''. Supported in part by the Deutsche Forschungsgemeinschaft (DFG)
under contract no.\ Re856/6-1,2.}} 

\vspace*{6mm}
{H. Reinhardt, D. Campagnari, D. Epple, M. Leder, M. Pak and W.
Schleifenbaum}\\      
{\small \it University of T\"ubingen \\      
            Institute of Theoretical Physics\\
	    Auf der Morgenstelle 14\\
	    D-72076 T\"ubingen}
\end{center}

\vspace*{6mm}

\begin{abstract}
Within the Hamiltonian approach  in Coulomb gauge the ghost and gluon
propagators are determined from a variational solution of the
Yang--Mills Schr\"odinger equation  showing both
gluon and heavy quark confinement. The continuum results  are in good agreement
with lattice data. The ghost form factor is identified as the
dielectric function of the Yang--Mills vacuum and a connection between
the Gribov--Zwanziger scenario and the dual Meissner effect is
established. The topological susceptibility is calculated.

\end{abstract}

\vspace*{6mm}

\section{Introduction}
The aim of the talk is the microscopic description of infrared properties of QCD
like confinement. We would like to see, for example, the emergence of the colour
flux string between static colour charges. For this purpose, I will use the
Hamiltonian approach to Yang-Mills theory in Coulomb gauge. The organisation of
my talk is as follows: In section 2, I will briefly summarise the essential ingredients of the
Hamiltonian approach to Yang-Mills  theory in Coulomb gauge. Then I will present
a variational solution of the Yang-Mills Schr\"odinger equation in
section 3, which will
result in a set of coupled Dyson-Schwinger equations. I will present the
analytic solutions to these equations in the infrared and ultraviolet and the
numerical solution for the full momentum range. The resulting propagators will
then be compared to the available lattice data. Then I will focus on two
non-perturbative properties of the Yang-Mills vacuum: the dielectric
constant in section 4 and
the topological susceptibility in section 5. Finally, a summary is provided.
\section{Canonical quantisation of Yang-Mills theory}
In the canonical quantisation approach the gauge fields $A^a_\mu (x)$ are
considered as the (cartesian) coordinates and the corresponding conjugate
momenta are defined by
\be
\label{G1}
\Pi^a_\mu (x) = \frac{\delta S}{\delta \partial_0 A^a_\mu (x)} \hk ,
\ee
where $S$ is the action of the Yang-Mills field. The explicit calculation yields
\be
\label{G2}
\Pi^a_i (x) = E^a_i (x) \hk , \hk \Pi^a_0 (x) = 0 \hk ,
\ee
where $E^a_i (x)$ is the colour electric field. To avoid the problems arising
from the vanishing temporal component of the canonical momentum, one
imposes Weyl gauge $A^a_0 (x) = 0$. The Yang-Mills Hamiltonian is then given by
\be
\label{G3}
H = \frac{1}{2} \int d^3 x \lk \vec{\Pi}^2 (x) + \vB^2 (x) \rk \hk .
\ee
The canonical quantisation is carried out in the standard fashion by imposing
the canonical commutation relation $[ A^a_i (x), \Pi^b_j (y) ] =
\delta_{ij} \delta^{a b} \delta^3 (x - y)$, which promotes the canonical
momentum to the operator $\Pi^a_i (x) = \delta / i \delta A^a_i (x)$. By
imposing Weyl gauge one loses Gauss' law from the Heisenberg equation of
motion and Gauss' law has to be imposed as a constraint on the wave functional
\be
\label{G4}
{\bf \hat{D}\cdot \Pi} (x) \psi (x) = -g\rho_m (x) \psi (x) \hk ,
\ee
where $\rho_m (x)$ is the colour charge density of the matter fields and
$\hat{D}^{a b}_i = \delta^{a b} \partial_i + g \hat{A}^{a b}_i \hk\;$ \mbox{$(\hat{A}^{a b} = f^{a c b} A^c$)} is the covariant derivative in the adjoint
representation of the gauge field with $f^{a b c}$ being the structure constant
of the gauge group. The operator on the left hand side of Gauss' law is nothing
but the generator of time-independent gauge transformations and in the absence
of external colour charges, $\rho_m (x) = 0$, Gauss' law expresses the
invariance of the wave functional under space-dependent but time-independent
gauge transformations. 

Instead of working with explicit gauge invariant wave
functionals it is more convenient to explicitly resolve Gauss' law by fixing
the gauge. Coulomb gauge is  a particular convenient choice for this purpose.
We implement the Coulomb gauge, $\vec{\partial}\cdot \vA = 0$, in the standard fashion
into the scalar product of the wave functionals by means of the Faddeev-Popov
method
\be
\label{G6}
\langle \psi | {\cal{O}} | \phi \rangle = \int D A^\perp J (A^\perp) \psi^* (A^\perp) {\cal{O}} [A^\perp]
\phi (A^\perp) \hk ,
\ee
where 
\be
\label{G7}
J = \mathrm{Det} (- \vec{\hat{D}} \cdot \vec{\partial} )
\ee
is the Faddeev-Popov determinant. While in Coulomb gauge the gauge field is
transversal the momentum operator $\vec{\Pi} = \vec{\Pi}^{||} + \vec{\Pi}^\perp$
 contains both longitudinal $\vec{\Pi}^{||}$ and
transversal $\vec{\Pi}^\perp$ 
parts. Resolving Gauss' law for the longitudinal part of the
momentum operator yields
\be
\label{G9}
\vec{\Pi}^{||}
 \psi = g\vec{\partial} (- \vec{\hat D}\cdot \vec{\partial})^{- 1} \rho\, \psi
\hk , \hk \rho = \rho_g + \rho_{m} \hk ,
\ee
where
\be
\label{G10}
\rho_g = {\bf \hat{A}^\perp\cdot \Pi^{\perp}}
\ee
is the colour charge density of the gauge field. With this result the
Hamiltonian in Coulomb gauge is found to be 
\be
\label{G11}
H = \frac{1}{2} \int \lk J^{- 1} \vec{\Pi}^\perp J \vec{\Pi}^\perp + \vB^2 \rk +
H_C \hk ,
\ee
where
\bea
\label{G12}
H_C & = & \frac{1}{2} \int J^{- 1} \vec{\Pi}^\parallel J \vec{\Pi}^\parallel  = \frac{g^2}{2} \int J^{- 1} \rho (- \vec{\hat{D}}\cdot \vec{\partial})^{- 1} (-
\vec{\partial}^2) (  - \vec{\hat{D}} \cdot\vec{\partial})^{- 1} J \rho \hk 
\eea
is the so-called Coulomb Hamiltonian, which arises from the longitudinal part of
the kinetic energy after resolving Gauss' law. The Hamiltonian (\ref{G11}) was
first derived in Ref.\ \cite{ChrLee80}.
\section{Variational solution}
\begin{figure}
  \centering
  \includegraphics{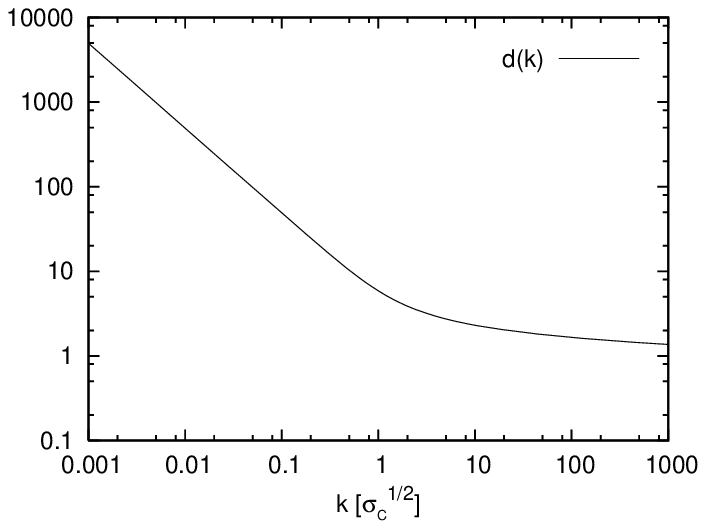}
  \includegraphics{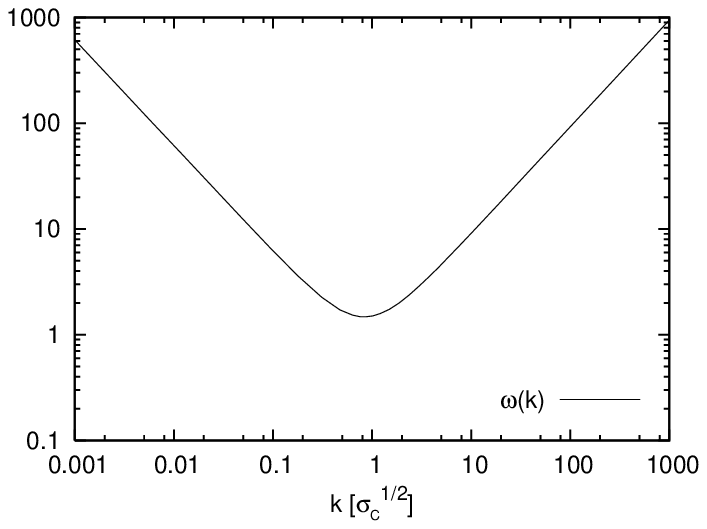}
  \caption{Ghost form factor $d(k)$ (left) and gluon energy $\omega(k)$ from the variational solutions presented in \cite{EppReiSch07}.}
  \label{d+w}
\end{figure}
\begin{figure}
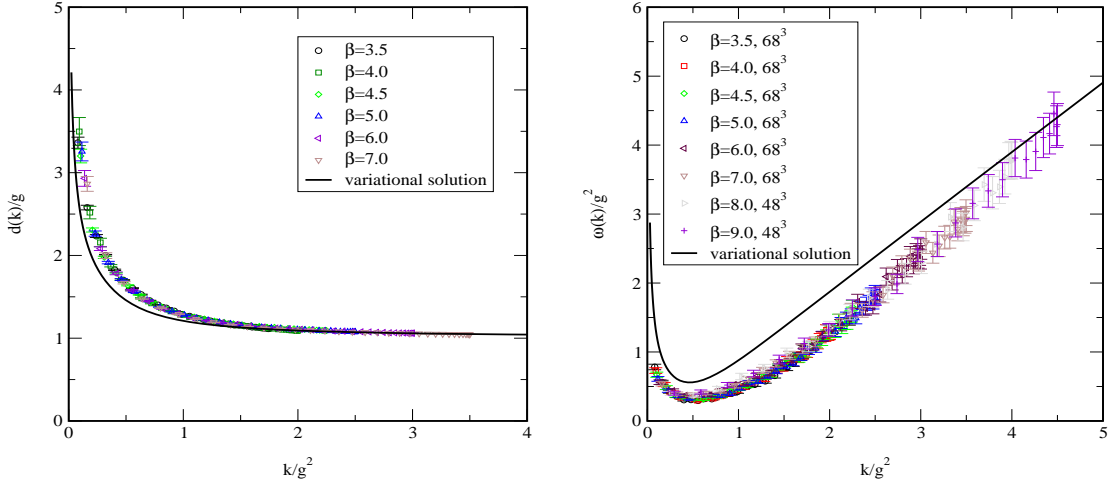

  \centering
  \includegraphics[width=7cm,height=6.4cm]{ghost_lin}\hspace{0.5cm}
  \includegraphics[width=7cm,height=6.4cm]{gapfunc_lin}
  \caption{Results for the ghost form factor $d(k)$ (left) and the gluon energy $\omega(k)$ in $2+1$ dimensions, as shown in \cite{FeuRei07}. }
  \label{2+1}
\end{figure}

We wish to solve the Schr\"odinger equation
\be
\label{G13}
H \psi = E \psi
\ee
by the variational principle
\be
\label{G14}
\langle \psi | H | \psi \rangle \to min \hk 
\ee
with suitable ans\"atze for the wave functional $\psi (A^\perp)$. This
approach has been studied resently in Refs.\ \cite{SzcSwa02,FeuRei04}. Inspired by the
wave functional of a massless particle moving in a spherically symmetric potential
in an $s$-state $\psi = \phi (r) / r$, where $r = {J^{1/2} (r)}$ is the Jacobian
of the transformation from the cartesian to the spherical coordinates for zero
angular momentum we choose the following ansatz \cite{FeuRei04}
\be
\label{G15}
\psi (A^\perp) = \frac{1}{\sqrt{J (A^\perp)}} \exp \lk - \frac{1}{2} \int d^3 x
d^3 y A^{\perp a}_i (x) \omega (x, y) A^{\perp a}_i (y) \rk \hk ,
\ee
where the kernel $\omega (x, y)$ is determined from the variational
principle (\ref{G14}). In practice the so resulting 
equation for $\omega (x, y)$ is
converted into a set of Dyson-Schwinger equations for the gluon propagator
\be
\label{G16}
\langle A^{\perp a}_i (x) A^{\perp b}_j (y) \rangle = \delta^{a b} t_{i j}(x) \frac{1}{2}
\omega^{- 1} (x, y) \; ,
\ee
with $t_{ij}(x)=\delta_{ij}-\frac{\partial_i\partial_j}{\partial^2}$
being the transverse projector,
and the ghost propagator
\be
\label{G17}
G (x, y) = \left\langle \lk - \vec{\hat{D}}\cdot \vec{\partial} \rk^{-1} \right\rangle
=  \langle x |d (-
\Delta) (- \Delta)^{- 1} | y \rangle \hk .
\ee
Here we have introduced the ghost form factor $d (- \Delta)$, which describes
the deviation of the QCD ghost propagator from the QED case, where $d (- \Delta)
\equiv 1$. The resulting Dyson-Schwinger equations need renormalisation, which
is well under control. Fig.\ \ref{d+w} shows the solution of the Dyson-Schwinger equation
for the gluon energy $\omega (k)$ and the ghost form factor $d (k)$,
as shown in Ref.\ \cite{EppReiSch07}. 
An analytic
infrared and ultraviolet analysis of the Dyson-Schwinger equation shows the
following asymptotic behaviour \cite{FeuRei04,SchLedRei06}
\bea
\label{G18}
\mathrm{IR}\, (k \to 0) & : & \omega (k) \sim \frac{1}{k} \hspace{1cm} d (k) \sim
\frac{1}{k} \nonumber\\
\mathrm{UV}\, (k \to \infty) & : & \omega (k) \sim {k} \hspace{1cm} d (k) \sim k^0
\hk .
\eea
At large momenta the gluon behaves like a photon, which is in agreement with
asymptotic freedom, while at small momenta the gluon energy diverges, which
implies the absence of gluon states in the physical spectrum. This is nothing
but a manifestation of gluon confinement. The infrared divergence of the ghost
form factor is a consequence of the horizon condition 
\be
\label{G19}
d^{- 1} (k = 0) = 0  \hk ,
\ee
which has been used as input in the renormalisation of the ghost Dyson-Schwinger
equation. This is a necessary condition for the Gribov-Zwanziger confinement
scenario. In fact, one can show there is a sum rule relating the infrared
exponents of the ghost and the gluon propagator and an infrared divergent gluon
energy requires also an infrared divergent ghost form factor, i.e.\ the horizon
condition (\ref{G19}), see Ref.\ \cite{SchLedRei06}. 
\begin{figure}
  \centering
  \includegraphics{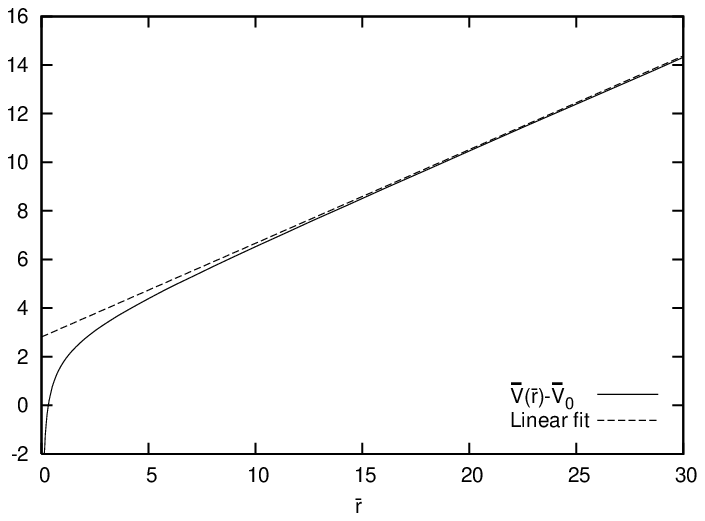}
  \includegraphics{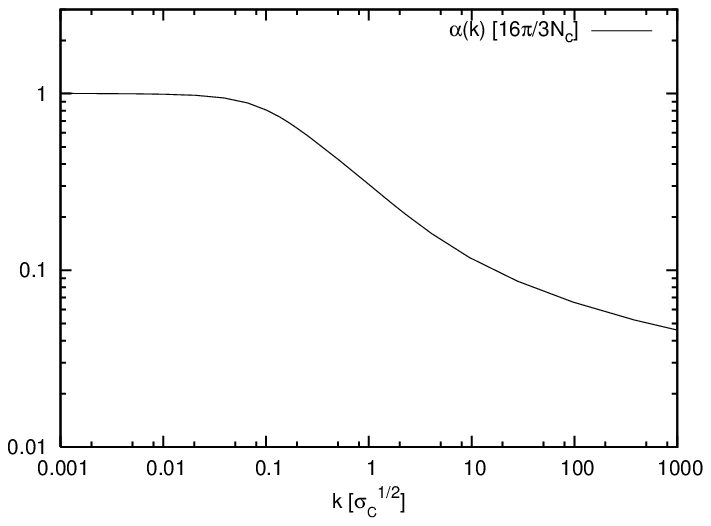}
  \caption{{\it Left:} Heavy quark potential given by eq.\
    (\ref{G20}). {\it Right:} Running coupling constant.}
  \label{Vc+alpha}
\end{figure}
Fig.\ \ref{Vc+alpha} shows the non-Abelian Coulomb potential
\be
\label{G20}
V (| x - y|) = g^2 \left\langle \langle x |
 ( - \vec{\hat{D}}\cdot \vec{\partial} )^{-1}
  (- \vec{\partial}^2) (- \vec{\hat{D}}\cdot \vec{\partial})^{- 1} | y
  \rangle \right\rangle \rightarrow \sigma_C\, |x-y|
\hk ,
\ee
which for large distance indeed increases linearly \cite{EppReiSch07} as the infrared analysis
reveals. The Coulomb string tension $\sigma_C$ sets the scale of our
approach. Also shown in Fig.\ \ref{Vc+alpha} is the running coupling constant which
is infrared finite, for details see Ref.\ \cite{SchLedRei06}.
Fig.\ \ref{2+1} shows the continuum results for the gluon energy and the ghost form factor in $D =
2 + 1$ dimensions \cite{FeuRei07} together with the corresponding
lattice results, Ref.\ \cite{diss_Moyaerts}. The agreement is not perfect
but, given the approximation involved, quite satisfactory. 

In $D = 3 + 1$ dimensions, previous lattice calculations performed in Coulomb gauge in
Ref.\ \cite{LanMoy04, Qua+07} showed an anomalous UV behaviour of the gluon
propagator --- $\mathrm{IR}: \omega (k) \sim k^0 \, ,\; \mathrm{UV}: \omega (k) \sim
k^{3/2}$ --- which is in strong conflict with the continuum result. However, one
should mention that these lattice calculations assumed multiplicative
renormalisability of the 4-dimensional gluon propagator, which give rise to
scaling violations in the static propagator. Furthermore, these calculations did
not fix the gauge completely, i.e.\ the residual time-dependent gauge invariance
left after Coulomb gauge fixing was left unfixed. Furthermore, the Coulomb
gauge fixing was done on a single time-slice, which is sufficient for the
calculation of static (time-independent) propagators. However, one should keep
in mind that in Coulomb gauge topologically non-trivial gauge configurations,
which presumably are responsible for confinement, are discontinuous in
time \cite{JacMuzReb78} and
as a consequence on a small lattice different results are obtained from
different time slices.

\begin{figure}
  \centering
  \includegraphics[scale=1.0]{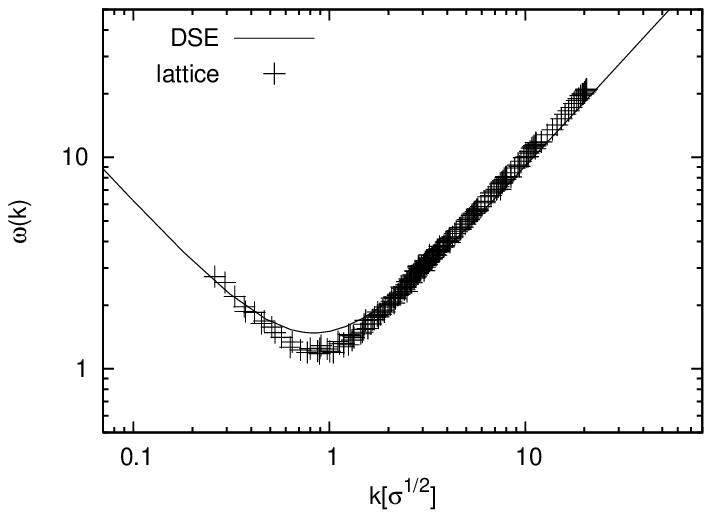}
  \includegraphics[scale=1.0]{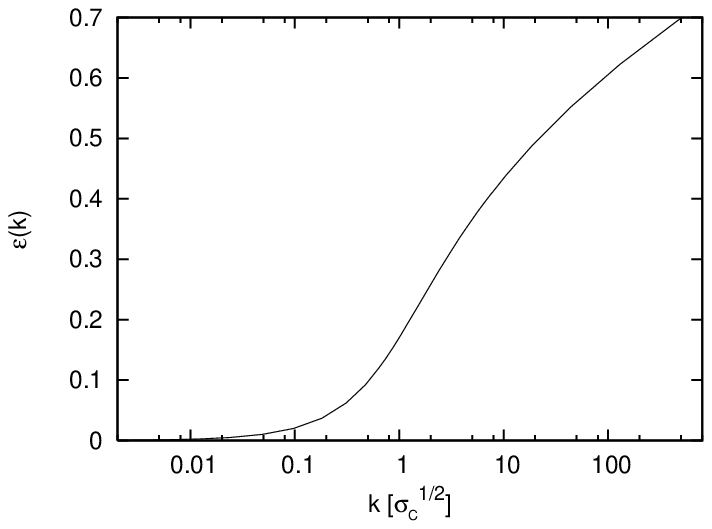}
  \caption{{\it Left:} Lattice data for $\omega(k)$, compared to the
    solution of   the Dyson-Schwinger equations.  {\it Right:} Dielectric function $\epsilon(k)$.}
  \label{latt+eps}
\end{figure}

Recently, we have done improved lattice calculations with a complete gauge
fixing \cite{Bur+08}. In these studies, the energy dependence of the 4-dimensional gluon propagator
could be explicitly extracted and it was found that the static gluon propagator
is multiplicatively renormalisable and shows a perfect scaling. 
Fig.\ \ref{latt+eps} (left panel) shows the
results for the gluon propagator of these calculations together with the
continuum results. It is assumed here that the Coulomb string tension
$\sigma_C$ is identical to the string tension $\sigma$ from the Wilson
loop. There is a very good agreement, in particular the ultraviolet
and infrared behaviour matches perfect for lattice and continuum. What is also
remarkable that the lattice result can be very well fitted by Gribov's original
formula for the gluon energy
\be
\label{21}
\omega (k) = \sqrt{k^2 + \frac{M^4}{k^2}}
\ee
with $M = 0.88(1)\,\mathrm{GeV}$.

\section{The colour dielectric constant}
Consider the electric field generated by a charge density $\rho$ in
electrodynamics
\be
\label{G21a}
{\bf E} = - \vec{\partial} \phi \hk , \hk \phi = (- \Delta)^{- 1} \rho \hk .
\ee
The longitudinal electric field resulting from the resolution of Gauss' law in
the Yang-Mills case is given by a similar expression
\bea
\label{G22}
{\bf E} & = & \langle {\bf \Pi} \rangle = - \partial \phi \hk \nonumber\\
\phi & = & \left\langle \lk - \vec{\hat{D}}\cdot \vec{\partial} \rk^{- 1} \right\rangle \rho = d
(- \Delta) (- \Delta)^{- 1} \rho \hk 
\eea
except that the Green's function of the Laplacian is replaced by the ghost
propagator (\ref{G17}). The last expression has the form of the scalar potential
in the presence of a dia-electric medium
\be
\label{G23}
\phi = \epsilon^{- 1} (- \Delta)^{- 1} \rho
\ee
and the inverse of the ghost form factor $d(k)$ can thus be identified as the dielectric
function of the Yang-Mills vacuum
\be
\label{G24}
\epsilon (k) = d^{- 1} (k) \hk .
\ee
Fig.\ \ref{latt+eps} (right panel) shows the so defined dielectric function. It satisfies $0 < \epsilon (k)
< 1$, which is a manifestation of anti-screening while in QED we have $\epsilon
(k) > 1$, which corresponds to ordinary Debye screening. Furthermore, at zero
momentum the dielectric function vanishes, showing that in the infrared the
Yang-Mills vacuum behaves like a perfect colour dia-electric medium. The vanishing of
the dielectric function in the infrared is not an artifact of our solutions of
the Dyson-Schwinger equations but is guaranteed by the horizon condition, which
is a necessary condition for the Gribov-Zwanziger confinement scenario. A
perfect colour dia-electric medium $\epsilon = 0$ is nothing but a dual
superconductor. (Here, ``dual'' refers to an interchange of electric and magnetic
fields and charges.) Recall in an ordinary superconductor the magnetic
permeability vanishes $\mu = 0$. This shows that the Gribov-Zwanziger
confinement scenario implies the dual Meissner effect \cite{Rei08}.

\begin{figure}
  \centering
  \includegraphics[scale=0.6]{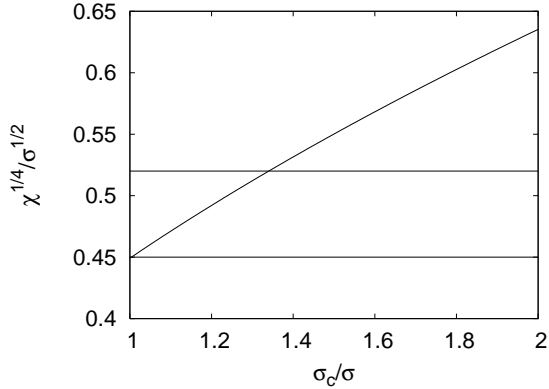}
  \caption{Topological susceptibility $\chi$ as a function of the ratio
 $\sigma_c /\sigma$.}
  \label{chi}
\end{figure}
\section{Topological susceptibility}
As first shown by Adler \cite{Adl69} and Bell and Jackiw \cite{BelJac69}, the $U_A (1)$
symmetry is anomalously broken which gives rise to an extra mass term to the
$\eta'$, which by the Witten-Veneziano formula
\be
\label{G25}
m^2_{\eta'} + m^2_\eta - 2 m^2_K = \frac{2 N_f}{F^2_\pi} \chi
\ee
is expressed by the topological susceptibility
\be
\label{G26}
\chi = - i \int d^4 x \langle 0 | q (x) q (0) | 0 \rangle \; ,
\ee
which is the correlation function of the topological charge density
\be
\label{G27}
q (x) = \frac{g^2}{32 \pi^2} F^a_{\mu \nu} (x) \tilde{F}^{a \mu \nu} (x) \hk .
\ee
Furthermore, in eq.\ (\ref{G25}) $N_f$ denotes the number of flavours and $F_\pi
\sim 93 \,\mathrm{MeV}$ is the pion decay constant. $\chi$ vanishes in all orders of
perturbation theory and is thus an ideal observable to test the non-perturbative
content of our vacuum wave functional. In the Hamiltonian approach one finds the
following expression for the topological susceptibility \cite{CamRei08}
\be
\label{G28}
V \chi = \lk \frac{g^2}{8 \pi^2} \rk^2 \left[ \langle 0 | \int \vB^2 (x) | 0
\rangle - 2 \sli_n \frac{| \langle n | \int \vB \cdot\vec{\Pi} | 0 \rangle |^2}{E_n}
\right] \hk .
\ee
Here $| n \rangle$ denotes the exact excited states of the Yang-Mills
Hamiltonian with energies $E_n$. These eigenstates are of course not known. We
work out the matrix elements in eq.\ (\ref{G28}) to two-loop order. In this order
only two and three quasi gluon states 
\be
a^{a^\dagger}_i (x) a^{b^\dagger}_j (y) | 0 \rangle \hk , \hk a^{a^\dagger}_i
(x) a^{b^\dagger}_j (y) a^{c^\dagger}_k (z) | 0 \rangle
\ee
contribute where our vacuum state is annihilated by the operators $a^a_i (x)$,
i.e.\ $a^a_i (x) | 0 \rangle = 0$. The resulting expression for the topological
susceptibility is ultraviolet divergent and needs renormalisation. For this aim
we exploit the fact that $\chi$ vanishes to all order perturbation theory and
renormalise the expression (\ref{G28}) for $\chi$
by subtracting each propagator by its perturbative
expression. This renders $\chi$ (\ref{G28}) finite. Furthermore, since the
momentum integrals in this expression are dominated by the infrared part we
replace the coupling constant, which, in principle, should be the running one,
 by
its infrared value. The results obtained in this way for the topological
susceptibility are shown in Fig.\ \ref{chi} (right panel) as a function of the ratio $\sigma_C / \sigma$. Choosing $\sigma_C = 1.5 \sigma$ 
which is the value favoured
by the lattice calculation \cite{LanMoy04} we find with
$\sqrt{\sigma} = 440 \hk MeV$
\be
\label{G30}
\chi = (240 \hk  MeV)^4 \hk .
\ee
This value is somewhat larger than the lattice prediction $\chi = (200 - 230 \hk 
MeV)^4$.
\section{Summary and Conclusions}
I have presented a variational solution of the Yang-Mills Schr\"odinger equation
in Coulomb gauge using a Gaussian type of ansatz for the vacuum wave functional.
We find a gluon energy which is infrared divergent, which is a manifestation of
gluon confinement. Furthermore, we have found a static colour charge potential which at
large distances rises  linearly, as one expects for a confining
theory. The propagators calculated within this approach are all in satisfactory
agreement with the lattice data. I have then shown that the inverse of the ghost
form factor can be interpreted as the colour dielectric function of the QCD
vacuum. The horizon condition,  a necessary condition for the Gribov-Zwanziger
confinement scenario to work, implies that in the infrared the QCD vacuum is a
perfect colour dia-electric medium, which is nothing but a dual superconductor.
In this way the Gribov-Zwanziger confinement scenario implies the dual Meissner
effect. Finally I have presented results for the topological susceptibility
calculated in the Hamiltonian approach with our vacuum wave functional. For
reasonable values of the Coulomb string tension we find results close to but
somewhat larger than the lattice data. The results obtained so far in this
approach are quite encouraging for further investigations. A natural next step
would be the inclusion of dynamical quarks.



\begin{thebibliography}{10}
\addtolength{\itemsep}{-6pt}
\bibitem{ChrLee80}
N.~H. Christ and T.~D. Lee,
\newblock Phys. Rev. {\bf D22}, 939 (1980).

\bibitem{EppReiSch07}
D.~Epple, H.~Reinhardt, and W.~Schleifenbaum,
\newblock Phys. Rev. {\bf D75}, 045011 (2007), hep-th/0612241.

\bibitem{SzcSwa02}
A.~P. Szczepaniak and E.~S. Swanson,
\newblock Phys. Rev. {\bf D65}, 025012 (2001), hep-ph/0107078.

\bibitem{FeuRei04}
C.~Feuchter and H.~Reinhardt,
\newblock Phys. Rev. {\bf D70}, 105021 (2004), hep-th/0408236.

\bibitem{SchLedRei06}
W.~Schleifenbaum, M.~Leder, and H.~Reinhardt,
\newblock Phys. Rev. {\bf D73}, 125019 (2006), hep-th/0605115.

\bibitem{FeuRei07}
C.~Feuchter and H.~Reinhardt,
\newblock Phys. Rev. {\bf D77}, 085023 (2008), 0711.2452.

\bibitem{diss_Moyaerts}
L.~Moyaerts,
\newblock {\em A numerical study of quantum forces},
\newblock PhD thesis, Univ. of T\"ubingen, Germany, 2004.

\bibitem{LanMoy04}
K.~Langfeld and L.~Moyaerts,
\newblock Phys. Rev. {\bf D70}, 074507 (2004), hep-lat/0406024.

\bibitem{Qua+07}
M.~Quandt, G.~Burgio, S.~Chimchinda, and H.~Reinhardt,
\newblock PoS {\bf LAT2007}, 325 (2007), arXiv:0710.0549 [hep-lat].

\bibitem{JacMuzReb78}
R.~Jackiw, I.~Muzinich, and C.~Rebbi,
\newblock Phys. Rev. {\bf D17}, 1576 (1978).

\bibitem{Bur+08}
G.~Burgio, M.~Quandt, and H.~Reinhardt,
\newblock (2008), 0807.3291.

\bibitem{Rei08}
H.~Reinhardt,
\newblock (2008), 0803.0504,
\newblock Phys. Rev. Lett., in press.

\bibitem{Adl69}
S.~L. Adler,
\newblock Phys. Rev. {\bf 177}, 2426 (1969).

\bibitem{BelJac69}
J.~S. Bell and R.~Jackiw,
\newblock Nuovo Cim. {\bf A60}, 47 (1969).

\bibitem{CamRei08}
D.~R. Campagnari and H.~Reinhardt,
\newblock (2008), 0807.1195.

\end{thebibliography}
\end{document}